\documentstyle[12pt]{article}
\def\beq{\begin{equation}}
\def\eeq{\end{equation}}
\def\ben{\begin{eqnarray}}
\def\een{\end{eqnarray}}
\def\bea{\begin{array}}
\def\eea{\end{array}}

\begin{document}

\baselineskip=18pt


\begin{flushright}
Prairie View A \& M, HEP-5-95\\
May 1995

\end{flushright}

\vskip.1in

\begin{center}
{\Large\bf Cascade Global Symmetry and Quark Mass Hierarchy
 }\\
{\small (a contribution to the Workshop on Part. Theo. \& Phenomenology,
 May 17-19, 1995, Ames, Iowa)}

\vskip .3in

{\bf Dan-di Wu}

{\sl HEP, Prairie View A\&M University, Prairie View, TX 77446-0355, USA}\\
{\small  $~$E-mail:  wu@hp75.pvamu.edu or DANWU@
physics.rice.edu }

\vskip .2in

\end{center}

\begin{abstract}

  Cascade global symmetry and multi-vacuum expectation values are combined to
produce  an {\bf initial} texture of the quark mass matrices.
Required corrections to the initial texture zeros ({\bf ITZ)} are all at the
order of
$10^{-3}m_t$ or less.
The possibility of {\bf radiative corrections} as the source of the
complete mass
matrices is briefly discussed.
\end{abstract}
\oddsidemargin.1in


The discovery of the top quark makes the fermion spectrum in the standard
 model complete. We are now puzzled by the pattern of the mass spectrum
in the standard model (SM)
 (Fig.1), as we were thirty years ago by the spectrum of hadrons. However,
 history does not repeat itself without some varieties. Obviously, the
 masses of quarks and leptons are regulated by mass ratios (hierarchy),
 instead of mass differences in the case of hadrons.
\vskip.05in

Dr. Tanaka has just reviewed some previous works in resolving this puzzle. I
 here would like to present an alternative approach which relates this puzzle
 to the structure of the Higgs sector, a sector yet completely in the dark.
So far, no electroweak scalar (Higgs) has ever been found. However, the best
 limits are only set for the minimal standard model Higgs boson. For a more
 complicated spectrum of scalar particles ({\bf why} are there
 not many scalars, while
there are so many fermions ?), only one mass eigenstate, which is
 in the direction of the physical vacuum, has dimension three
 couplings with the weak gauge bosons; all the other Higgs particles, if they
 exist,
 are orthogonal to this single state and possess only dimension four couplings
 with gauge bosons, and therefore are more difficult to discover. The main
feature
 of this approach is to use a natural global symmetry of the standard
gauge-fermion
interaction sector to {\bf protect} the small matrix elements in the
 mass matrices of the quarks.
\vskip.05in

The basic ideas behind this approach[1,2] are  the following:\\
1.$~$Approximate global symmetry which is non-abelian in order to be more
 restrictive.\\
2.$~$Multi-Higgs doublets with different vacuum expectation values (VEVs)
contribute
 to dif-

 ferent mass terms.\\
3.$~$The naturalness principle which is engaged to control the sizes of the
coupling
  constants

  whose
ratios are at the order of 10$^0$.\\
4.$~$Rich scalar spectrum with a bunch of pseudo-Nambu-Goldstone bosons
 (PNGB).
\vskip.05in

Approximate global symmetries have been found in strong interactions. I-spin,
 chiral
$SU(3)\times SU(3)$ and G-parity, to name just a few. These global
 symmetries are inexact. In the example of $SU(3)\times SU(3)$ chiral
 symmetry, it is a common belief that  symmetry breaking comes from both
 the existence of a non-trivial QCD vacuum and the existence of the current
 quark masses. The former is called spontaneous symmetry breaking (SSB);
 and the latter, explicit symmetry breaking (ESB). A pattern was emerged
 from the relevant studies: While gauge symmetries are regarded as exact and
 dynamical symmetries, global symmetries are regarded as inexact and
accidental.
 ESB of global symmetries are assumed ad hoc  in the effective Lagrangain,
although the ESB itself may be the effect of another SSB at a higher scale,
as in
the case of current masses of quarks in the chiral theory of light hadrons.
\vskip.05in

The specific global symmetry we  use is ${\bf G}=SU(3)\times U(1)$, where
 $3$ corresponds to three generations and the extra $U(1)$ is  an overall
 phase transition. Left-handed quarks are in a triplet (3), in addition to
being
 in $SU(2)_L$ doublets. Right-handed quarks are in
two anti-triplets (3$^*$). The global $U(1)$ quantum number is also assigned
 differently for the left and right handed fields. A Higgs triplet and a Higgs
sextet,
which are also weak doublets,
are introduced  to provide two different vacuum expectation
 values. These VEVs spontaneously break the ${\bf G}$ symmetry down to a
 $U(1)$, because in the chosen parameter space the two VEVs are
 perpendicular to each other.
\vskip.05in

It should be emphasized that the Yukawa sector does not respect exact global
 symmetry. There are  terms  with smaller
symmetries. These terms partially break the global symmetry. Since
the ESB terms introduced in the Yukawa sector are dimension-4
 operators, induced symmetry breaking terms in the Higgs potential are
divergent (Fig. 2). Therefore corresponding ESB must also be
introduced  in the Higgs sector in order to provide counter terms to these
 divergences.
\vskip.05in

Typically, the mass matrices produced directly by the Yukawa sector through SSB
are only the main texture. For example,
the mass matrix for the up type quarks is
\ben
M^U=\left(\bea{ccc}
0&0&0\\
	             0&0&G_1{\it v}^{\prime }\\
	              0&-G_1{\it v}^\prime &G_0{\it v}\eea\right).
\een
That for down type quarks is
\ben
M^D=\left(\bea{ccc}
0&0&0\\
	              0&0&G_3{\it v}^{\prime }\\
	              0&-G_3{\it v}^\prime &G_2{\it v}\eea\right).
\een
Here ${\it v}$ is the VEV of the sextet Higgs and ${\it v}^\prime$ is the VEV
of
the triplet Higgs. We assume ${\it v:v}^\prime\sim
1:0.4$. $G_\alpha$ with $\alpha=0,1,2,3$ are the Yukawa coupling constants
 which share sequentially smaller symmetries.  Approximately, the ratios of
  these couplings are
\beq
  G_0:G_1\sim 5, \hskip .2in G_1:G_2\sim 7, \hskip.2in G_2:G_3\sim 2.5 .
\eeq
Both the mass matrices in Eqs. (1) and (2) are rank-2 matrices and it is easy
to
 see that $m_t= |G_ov|$, $m_b=|G_1 v|$  and the masses $m_c$ and $m_s$
are respectively decided by a ``see-saw" like mechanism, which gives
$m_c=|G_1{\it v}^\prime|^2/m_t,\ \ m_s=|G_3{\it v}^\prime|^2/m_b$.
We also immediately obtain
$V_{cb}=x-x^\prime,$ with  $x=G_3{\it v^\prime}/G_2{\it v},\ \
 x^\prime =G_1{\it v^\prime}/G_0 {\it v}$.
This $V_{cb}$ is within a factor of two close to the experimental data in
 magnitude. Considering the preliminary property of this calculation,
 this result
is  encouraging. Impressive  progress has been made
 along other lines of thinking (to name just a few, see Refs.[3-7]).
However, new features of this approach will immediately become apparent.
\vskip.05in

A few other  interesting features of these two mass matrices should be noticed.
First, there are only four different entries of sequentially smaller
 magnitudes. The sequence is caused by both smaller VEV and smaller
 Yukawa coupling constants. Second, each mass matrix has six initial
 texture zeros (ITZs).
The nonzero matrix elements (2,3) or (3,2) in each matrix are about 10
 times smaller than the corresponding (3,3) elements. The required corrections
to all the ITZs are at the order of $10^{-3} m_t $ or less.
Third, the ITZs are protected by the symmetrical property of the Yukawa
 sector. All lowest loop corrections to them are either vanishing or finite.
Because the scalars which mediate flavor changed neutral (or charged)
 currents carry $SU(3)\times U(1)$ charges (called global charges), the
 loop will not be able to be closed, unless a mixing of scalars with
 different global charges is introduced. Because of these features,
 this approach may be called the {\bf ITZ}.
\vskip.05in

  This approach
  is  different from other approaches in that it
differentiates ITZs from non ITZs, protects them, and has a potential
to correct the ITZs by radiative
  corrections[8]. It also predicts many new scalar mediated processes, such as
  $t \rightarrow c c \bar u$, and corrects the rates of
  many processes of the standard model. In addition, this approach, if
successful,
 may change our
concept on the {\bf desert} between $10^4$ to $10^{15}$ GeV .
\vskip.05in

The following issues with respect to  this approach are under consideration.

{\bf A}. It is possible that the features of the initial mass matrices in
 Eqs.(1) and (2) allow one to produce (2,2),  (1,2) and (2,1) elements and
others, if necessary, by radiative
 corrections, in particular because the coupling constant  $G_0$ is at
  order 1. Therefore the correct masses of the first generation and the
whole mixing
 matrix (the CKM matrix) are calculable in some sense. This calculation may
 provide a method to restrict the
parameters of the ESB terms, at least those which play significant
roles in the calculations. With  luck one may even over constrain
these parameters. One is then in  good shape to calculate corrections to
 e.g. $B_s - \bar B_s$ mixing, $b\rightarrow s + \gamma$ and other measurable
 processes.  (See Fig. 3 for typical scalar mediated
 processes, where crosses on the dotted lines represent mixing vertices.)
 It is worth noting that because of the restriction of the
 $SU(2)\times U(1)$ gauge symmetry, it is impossible to introduce the needed
 mixing for loop diagrams which correct ITZs in the original Lagrangian.
Therefore this
 mixing must be
 a combined effect of ESB and SSB.
 \vskip.05in
{\bf B}. The Higgs potential sector. In addition to the symmetric $SU(3)\times
U(1)$
part, there must be partially symmetric and completely asymmetric
terms in  the Higgs potential. As discussed
 previously, the complete symmetry will be broken by the VEVs of the triplet
 and sextet Higgs down to $U(1)$. This SSB will cause 8+3 massless
 Nambu-Goldstone bosons. Three of these massless bosons which are
 related to gauge symmetry breaking are absorbed by the $W$ and $Z$
 bosons.
In order to transmute the other 8  massless
 bosons, which are related to the global symmetry breakdown, into massive ones,
one must  introduce ESB terms in the Higgs potential. ESB
 terms are necessary also to keep the theory consistent and
renormalizable (Fig. 2).
A third purpose of introducing ESB terms is to
 allow global charge breaking processes to occur, as discussed before.
For these reasons, the Higgs sector could be complicated.
In particular, there are many different
 ways to introduce ESB terms. A clever guideline for introducing ESB
terms may drastically simplify the solution. As emphasized in {\bf A}, much
 physics can be discussed before completely resolving the Higgs sector.
 \vskip.05in
{\bf C}. If the ESB terms in the Higgs sector are small, compared with the
 symmetric terms, then the masses of the formerly Nambu-
Goldstone bosons  will be small, compared with the masses in the
 original Higgs sector before SSB. Therefore there will be 8
 psuedo-Nambu-Goldstone bosons (PNGB). Because these particles are relatively
light, low energy physics is related to them. The properties
 of PNGBs are therefore worth studying.
 For example, the global quantum numbers of the PNGBs should be
 identified and the mass spectrum should be given in accordance with the
 pattern of the ESB.
\vskip.1in

In any case, the separation of the non-zero elements
in Eqs.(1) and (2)  from other small elements will shed new light to produce
 a mass and mixing pattern
 which is sufficiently hierarchical.

\vskip.1in

\noindent {\large\bf References: }\\
1. D. D. Wu, PVAMU-3-95, LANL bulletin: hep-ph/9504381.\\
2. S. Weinberg, UTTG-05-92, unpublished.\\
3. S. Weinberg, Transaction of the NY Acad. of Sci.,
Series II, Vol. 38, 185 (1977).\\
4. H. Fritzsch, Phys. Lett. B73 (1978) 317;
D. S. Du and Z. Z. Xing, Phys. Rev. D49 (1993)

2349; P. Ramond, R. G. Roberts,
and G. G. Ross, Nucl. Phys. 406 (1993) 19.\\
5. S. Babu and R. N. Mahapatra, Phys. Rev. Lett. 74 (1995) 2418 and therein.\\
6. See e.g. P. Framton and O. C. W. Kong, forthcoming in Phys. Rev. Lett.\\
7. C. D. Froggat and H.B. Nielson, Nucl. Phys. B147 (1979) 277; L. Ibanez and

G.G. Ross, Phys. Lett. B332 (1994) 100.\\
8. For radiative correction as a method to produce small elements in a  broad
context, see,

W. Marciano, in the Top Physics session of these proceedings.

\end{document}